\documentclass[11pt,twocolumn,a4paper,twoside]{article}

\usepackage{booktabs} % For formal tables
\usepackage[english]{babel}
\usepackage{amsmath,amssymb,amsfonts}
\usepackage{algorithmic}
\usepackage[english,ruled]{algorithm2e}
\usepackage{graphicx}
\usepackage{textcomp}
\usepackage{xcolor}

\usepackage{listings}

\date{October 2019: Revised Version}

\usepackage[caption=false,font=footnotesize,labelfont=sf,textfont=sf]{subfig}
\usepackage{url}

\usepackage{tabularx}
\newcolumntype{L}[1]{>{\raggedright\arraybackslash}p{#1}}
\newcolumntype{C}[1]{>{\centering\arraybackslash}p{#1}}

\begin{document}

\title{Towards Enabling Novel Applications though Hybrid Cloud/Edge Architectures\thanks{The work presented here was partially supported by the Lightkone European H2020 Project (under grant number 732505) and NOVA LINCS (through the FC\&T grant UID/CEC/04516/2013).}}
\author{
{Jo{\~a}o Leit{\~a}o, Pedro {\'A}kos Costa, Maria Cec{\'i}lia Gomes, and Nuno Pregui{\c c}a}\\
\small jc.leitao@fct.unl.pt, pah.costa@campus.fct.unl.pt, \{mcg, nuno.preguica\}@fct.unl.pt\\
\small NOVA LINCS \& DI-FCT-UNL\\ [2mm]
%\small Submission Type: Vision
}

%\author{Paper \#1199}

\maketitle

\begin{abstract}
Edge computing has emerged as a distributed computing paradigm to overcome practical scalability limits of cloud computing. The main principle of edge computing is to leverage on computational resources outside of the cloud for performing computations closer to data sources, avoiding unnecessary data transfers to the cloud and enabling faster responses for clients.

While this paradigm has been successfully employed to improve response times in some contexts, mostly by having clients perform pre-processing and/or filtering of data, or by leveraging on distributed caching infrastructures, we argue that the combination of edge and cloud computing has the potential to enable novel applications. However, to do so, some significant research challenges have to be tackled by the computer science community. In this paper, we discuss different edge resources and their potential use, motivated by envisioned use cases. We then discuss concrete research challenges that are in the critical path towards realizing our edge vision. We conclude by proposing a research agenda to allow the full exploitation of the potential for the emerging hybrid cloud/edge paradigm.
\end{abstract}

%\let\section\newsection
%\let\subsection\newsubsection

%
% The code below should be generated by the tool at
% http://dl.acm.org/ccs.cfm
% Please copy and paste the code instead of the example below.
%

%\keywords{Edge Computing, Cloud Computing, Self-Adaptive Systems}

\section{Introduction}

Since its inception in 2005, cloud computing has deeply impacted how distributed applications are designed, implemented, and deployed. Cloud computing offers the illusion of infinite resources available in data centers, whose usage can be elastically adapted to meet the needs of applications. Furthermore, data centers in different geographical locations enable application operators to provide better quality of service and availability for large numbers of users scattered around the world through the use of cloud-based techniques, such as geo-distribution and geo-replication.

Cloud computing however, is not a panacea for building reliable, available, and efficient distributed systems. In particular, the increasing popularity of Internet of Things (IoT) and Internet of Everything (IoE) applications, combined with an increase in mobile and user-centric applications, has led to a significant increase in the quantity of data being produced by application clients. Although cloud computing infrastructures are highly scalable, the time required to process such large amounts of data is becoming prohibitively high. Additionally, the network capacity between clients and data centers is now becoming a significant bottleneck for such applications, namely to timely push data and fetch computation results to, and from the cloud. In fact, this has been recognized as a key contributing aspect for the increasing inability of cloud computing platforms to support data-intensive applications\,\cite{viewCloud}.

Due to this, moving computations towards the edge of systems (i.e., closer to the clients that effectively process and consume data) has become an essential endeavor to sustain the growth of such applications. This led to the emergence of \emph{edge computing}. Edge computing can be defined, in very broad terms, as performing computations outside the boundaries of data centers\,\cite{edge}. Many approaches have already leveraged on some form of edge computing to improve the latency perceived by end-users, such as CDNs\,\cite{akamainetsession}, or tapping into resources of client devices\,\cite{p2p,pplive}, among others.

This has motivated the emergence of different architecture proposals for taking advantage of edge computing. In particular, Cisco has proposed the model of Fog Computing\,\cite{cisco} which aims at improving the overall performance of IoT applications by placing servers (and network equipment with computing capacity) close to sensors that generate large amounts of data. These (Fog) servers can then pre-process data enabling timely reaction to variations on the sensed environment, and filter the relevant information that must be propagated towards cloud infrastructures for further processing.
Mist computing, is an evolution of the Fog computing model, that has been adopted by industry\,\cite{ibm} and that, in its essence, proposed to push computation towards sensors in IoT applications, enabling sensors themselves to perform data filtering computations, and alleviate the load imposed on Fog and Cloud servers. While these novel architectures exploit the potential of edge computing, they do so in a limited way, requiring specialized hardware and not taking a significant advantage of computational devices that already exist in the edge.
As noted for instance in~\cite{cisco}~and~\cite{ibm}, all of these architecture proposals are highly biased towards addressing the specific challenges found in IoT applications.

In this paper, we argue that edge computing also offers the opportunity to build new \emph{edge-enabled} applications, whose use of edge resources go beyond what has been done in the past, and in particular beyond proposals such as Fog and Mist computing\,\cite{cisco,ibm}. Previous authors have already presented their visions for the future of Edge computing\,\cite{edge,edge2}, Fog computing\,\cite{Vaquero:2014:FYW:2677046.2677052,Mahmud2018,8100873}, and IoT specific edge challenges\,\cite{7879243}.
These works however, present their visions with an emphasis on IoT applications. An exception to this is related with Mobile edge computing\,\cite{7879258} which devotes itself to the close cooperation of mobile devices to offload pressure from the cloud.
Contrary to these, we take a different approach on edge computing and envision a future where general user-centric applications will progressively grow to accommodate increasing numbers of users, while providing continuous access in an efficient way and enabling users to become more engaged in such applications. To achieve this, applications will evolve to leverage on a myriad of different and already existing edge resources, meaning that such applications will be (concurrently) executed across large number of heterogeneous components. Our vision, is that this will empower the design of user-centric applications that promote additional interactivity among users and between users and their (intelligent) environment.

\begin{figure*}[t]
{\center
\includegraphics[width=0.7\textwidth]{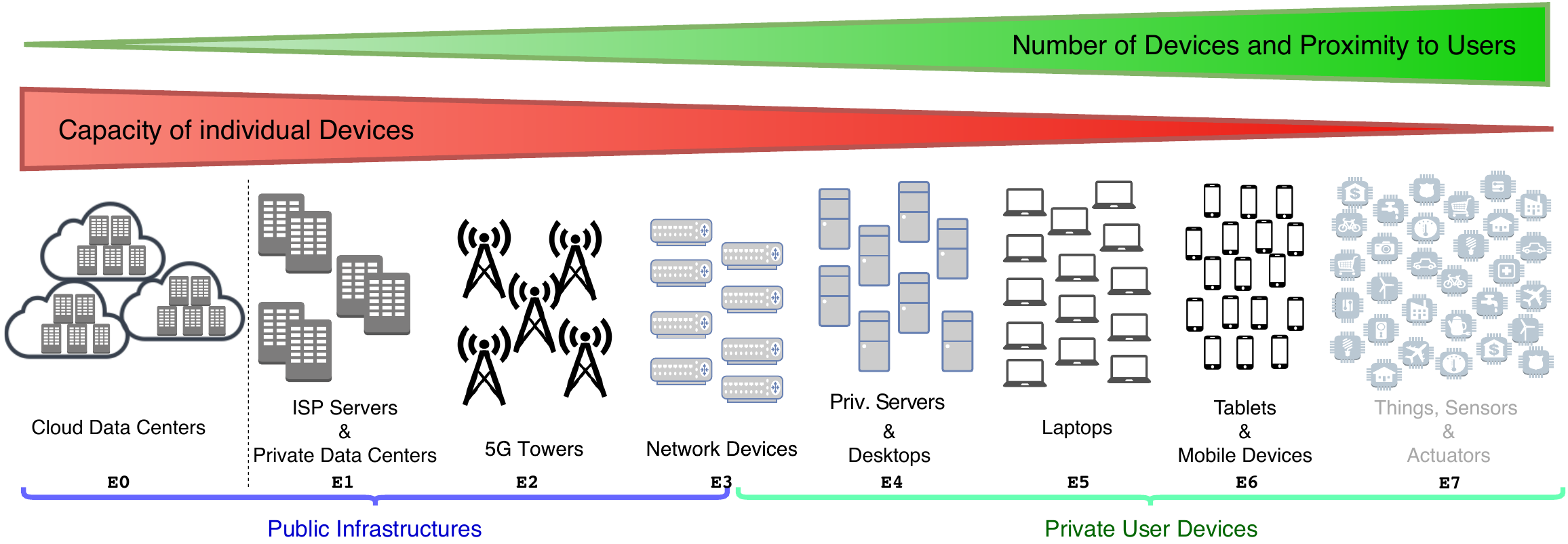}
\caption{Edge Components Spectrum\label{fig:edge}}
}
\end{figure*}

%To realize this vision however, it is relevant to fully understand what are the computational and network resources made available for edge-enabled applications. Furthermore, it is essential to understand how application developers and system architects can fully tap into these edge resources, what are the key technical and scientific challenges that have to be surpassed by the computer science community, and how these challenges are related with previous contributions.

%To overcome these needs and from a computer systems perspective, in this paper we make the following contributions. We start by presenting an overview of a general architecture for edge computing, identifying the different computational and network resources and explaining how they differ among each other ($\S$~\ref{sec:edge}). We further materialize our vision on the potential of edge computing by presenting two envisioned case studies ($\S$~\ref{sec:usecases}). We then identify concrete research challenges for enabling future edge-enabled applications and relate these with previous research to tackle similar challenges  ($\S$~\ref{sec:requirements}). Finally, we conclude this paper summarizing our proposed research agenda ($\S$~\ref{sec:conclusion}).

To realize this vision however, it is relevant to fully understand what are the computational and network resources available for edge-enabled applications, their characteristics, and how they can be used ($\S$\ref{sec:edge}). We further materialize our vision on the potential of edge computing by presenting two envisioned case studies ($\S$\ref{sec:usecases}). Using large numbers of heterogeneous edge resources to build novel edge-enabled applications is not trivial, and key research challenges must be addressed by the computer systems community, we further relate these with previous research ($\S$\ref{sec:requirements})\footnote{We recognize that fully tapping into the potential of edge computing requires concerted research efforts from many fields in computer science however, in this paper we focus on a computer systems perspective.}. Finally, we summarize, in the form of a research agenda towards the sustainable growth of edge computing architectures, the key open research challenges and some interesting departure points to fully realize our vision$\S$\ref{sec:agenda}).

%Budget in pages:
\section{Overview of the Edge}\label{sec:edge}

To fully realize the potential of edge computing, one should identify which computational resources lie beyond the cloud boundary, what are their limitations, and their potential benefits for edge-enabled applications. Figure~\ref{fig:edge} provides a visual representation of the different edge resources that we envision. We represent these edge resources as being organized in different levels starting with level zero that represents cloud data centers. Edge-enabled applications are not however, required to make use of resources across all edge levels.
%However, we expect data to mostly move between components that are adjacent in the spectrum. Levels can be skipped and different application data may follow different routes among edge components.

To better \emph{characterize} the different levels in the edge resource spectrum, we consider three main dimensions: $i)$ \textbf{capacity}, which refers to the processing power, storage capacity, and connectivity of the device; $ii)$ \textbf{availability}, which refers to the probability of the resource to be reachable (due to being continuously active or faults); and $iii)$ \textbf{domain}, which captures if the device supports the operation of an edge-enabled application as a whole (\emph{application domain}) or just the activities of a given user\footnote{We refer to user in broad terms, meaning an entity that uses an edge-enabled application, either an end-user or a company.} within an edge-enabled application (\emph{user domain}).

We further classify the \emph{potential uses} of the different edge resources considering two main dimensions: $i)$ \textbf{storage}, which refers to the ability of an edge resource to store and serve application data. Devices that can provide storage can do so by either storing \emph{full application state}, \emph{partial application state}, or by providing \emph{caching}. The first two enable state to be modified by that resource, and the later only enables reading (of potentially stale) data; and $ii)$ \textbf{computation}, which refers to the ability of performing data processing. Here, we consider three different classes of data processing, from the more general to the more restrictive: \emph{generic computations}, \emph{aggregation and summarization}, and \emph{data filtering}. We assume that an edge resource capable of performing the general classes of computations is also capable of performing the more restrictive ones.

We start by observing that as we move farther from the cloud (i.e, to higher edge levels), the capacity and availability of each individual resource tends to decrease, while the number of devices increases. We now discuss each of these edge resources in more detail. We further note that resources could be presented with different granularity however, in this paper we focus on a presentation that allows to distinguish computational resources in terms of their properties and potential uses within the scope of future edge-enabled applications.

\begin{table*}[t]
\begin{center}
{\scriptsize
\begin{tabular}{ll@{ }cccc}\toprule
&&{\bf E0}	&{\bf E1}	&{\bf E2}	&{\bf E3}\\
&&Cloud DCs&ISP Servers \& Priv. DCs&5G Towers&Network Devices\\
\midrule
						 			& {\bf Capacity} 	 		& High				& Large				& Medium					& Low\\
{\bf Characterization}	& {\bf Availability} 		& High				& High					& High						& High\\
									& {\bf Domain}		 		& Application 	& Application	   	& Application			& Application\\
\midrule
{\bf Potential uses}		& {\bf Storage}		 		& Full State		& (Large) Partial 	& (Limited) Partial	& None/Caching\\
									& {\bf Computation} 	& Generic			& Generic			   	& Generic					& Filtering	\\
\bottomrule
\end{tabular}

\vspace{0.5cm}

\begin{tabular}{ll@{ }cccc}\toprule
&&{\bf E4}	&{\bf E5}	&{\bf E6}	&{\bf E7}\\
&&Priv. Servers \& Desktops&Laptops&	Tablets \& Mobiles&Things\\
\midrule
						 			& {\bf Capacity} 	 		& Medium				& Medium			& Low					& Varied\\
{\bf Characterization}	& {\bf Availability} 			& Medium				& Low				& Low					& Limited\\
									& {\bf Domain}		 		& User					& User				& User					& User\\
\midrule
{\bf Potential uses}		& {\bf Storage}		 		& (User) Partial		& Caching			& (User) Caching 	& (Local) Caching\\
									& {\bf Computation} 		& Generic				& Aggregation	& Aggregation		& Filtering\\
\bottomrule
\end{tabular}
}
\end{center}
\caption{Edge Devices Per Level Characteristics\label{tab:edge}}
\end{table*}

\textbf{E0}: \emph{Cloud Data Centers.} Cloud data centers offer pools of computational and storage resources that can be dynamically scaled to support the operation of edge-enabled applications. The existence of geo-distributed locations can be used as a first edge computing level, by enabling data and computations to be performed at the data center closest to the client. These resources have \emph{high capacity and availability} and operate at the \emph{application domain}. They offer the possibility for storing \emph{full application state} and perform \emph{generic computations}.

\textbf{E1}: \emph{ISP Servers \& Private Data Centers.} This edge resource represents regional private data centers and dedicated servers located in Internet Service Providers (ISPs) or exchange points that can operate over data produced by users in a particular area. These servers operate at the \emph{application domain}, presenting \emph{large capacity} and \emph{high availability}. They offer the possibility to store \emph{(large) partial application state} and perform \emph{generic computations}.

\textbf{E2}: \emph{5G Towers}. The new advances in mobile networks will introduce processing and storage power in towers that serve as access points for mobile devices (and tablets) as well as improved connectivity. While we can expect these edge resources to have \emph{medium capacity}, they should have \emph{high availability} being able to operate at the \emph{application domain} particularly by serving users in their geographic areas of coverage. These computational resources can execute \emph{generic computations} over (potentially local) stored \emph{limited partial application state} enabling further interactions among clients (e.g, mobile devices) in close vicinity.

\textbf{E3}: \emph{Network Devices.} Network devices (such as routers, switches, and access points) that have processing power capabilities, offer \emph{low capacity} and \emph{high availability}. From the storage perspective, these offer either none or \emph{caching} capacity. Devices in close vicinity of the user will operate at the \emph{user domain} while equipment closer to the center of the system (i.e., lower edge levels) might operate at the \emph{application domain}. These devices will mostly enable in-network processing for edge-enabled applications in the form of \emph{data filtering} activities over data produced by client devices being shipped towards the center of the system.

\textbf{E4}: \emph{Private Servers \& Desktops.} This is the first level (and more powerful in terms of capacity) of devices operating exclusively in the \emph{user domain}. Private servers and desktop computers can easily operate as logical gateways to support the interaction and perform computations over data produced by levels E5-E7. While individually these edge resources have \emph{medium capacity} and \emph{medium availability} they can easily perform more sophisticated computing tasks if the resources of multiple devices are combined together. These edge resources are expected to store \emph{(user-specific) partial application state} while enabling \emph{generic computations} to be performed. Private servers in this context are equivalent to \emph{in-premises servers} frequently referred as part of Fog computing architectures\,\cite{cisco}.

\textbf{E5}: \emph{Laptops.} User laptops are similar to resources in the E4 level albeit, with \emph{low availability}. Low availability in this context is mostly related with the fact that the up-time of laptops is expected to be significantly lower due to the user moving from location to location. Due to this, we expect these devices to be used for performing \emph{aggregation and summarization} computations and eventually provide \emph{(user-specific) caching} of data for components running farther from the cloud. Laptops might act as application interaction portals, enabling users to use such devices to directly interact with edge-enabled applications. We note that, being an interaction portal for users, and due to their (potentially) limited connectivity, Laptops (and also edge resources in the E6 edge level discussed below) could significantly benefit from techniques to enable the use of applications in off-line mode, storing locally and later synchronizing (and in some cases validating) operations performed by the user while offline when connectivity is reestablished.

\textbf{E6}: \emph{Tablets \& Mobile Devices.} Tablets and Mobile devices are nowadays preferred interaction portals, enabling users to access and  interact with applications. We expect this trend to become dominant for new edge-enabled applications since users expect continuous and ubiquitous access to applications. These devices have \emph{low capacity} and \emph{low availability}, the latter is mostly justified by the fact that the battery life of these devices will shorten significantly if the device is used to perform continuous computations and data transfers to other edge levels. These devices however, can be used as logical gateways for devices in the E7 level in the \emph{user domain} context. They can provide \emph{user-specific caching} storage and perform either \emph{aggregation and summarization} or \emph{data filtering} for data produced by E7 devices in the context of a particular user.

\textbf{E7}: \emph{Things, Sensors, \& Actuators.} These are the most limited devices, in terms of capacity, in our edge resource spectrum. These devices will act in edge-enabled applications mostly as data producers and consumers. They have \emph{extremely limited capacity} and \emph{varied availability} (in some cases due to limited power and weak connectivity). They operate in the \emph{user domain}, and can only provide extremely limited forms of \emph{caching} for edge-enabled applications. Due to their limited processing power they are restricted to perform \emph{data filtering} computations. Devices in the E7 level with computational capacity are the basis for Mist computing architectures\,\cite{ibm}.

~\\

Table~\ref{tab:edge} summarizes the different characteristics and potential uses of edge resources at each of the considered levels. We expect application data to flow along the edge resource spectrum although, different data might be processed differently at each level (or skip some entirely).

%Budget in pages:
\section{Envisioned Case Studies}\label{sec:usecases}

We now briefly discuss two envisioned case studies of novel edge-enabled applications, and argue how edge resources in different levels of the edge spectrum can be leveraged to enable or improve these case studies.

\paragraph{Mobile Interactive Multiplayer Game} Consider an augmented reality mobile game that allows players to use their mobile devices to interact with augmented reality objects and non-playing characters similar to the popular P{\'o}kemon Go game\,\footnote{\url{https://www.pokemongo.com/}}. Such game could enable direct interactions among players, (e.g., to trade game objects or fight against each other) and allow players to interact in-game with (local) third party businesses that have agreements with the company operating the game (e.g., a coffee shop that offers in-game objects to people passing by their physical location).

P{\'o}kemon Go only recently started to enable some of these functionalities, and only in very limited forms. For instance, trades and battles can only be executed by players in close vicinity which are register as friends, being the latter poorly reactive to user input. There are some evidences \cite{pokemongolimitations} pointing to one of the main reasons being the inability of cloud-based servers to support such interactions in a timely manner due to large volumes of traffic produced by the application. However, edge computing offers the possibility to enable such interactions by leveraging on edge resources located in some of the levels discusses above.
Considering that the game is accessed primarily through mobile phones, one could resort to computational and storage capabilities of \emph{5G Towers (E2)} to mediate direct interactions (e.g., fights) between players. This would make interactions much more fluid, interactive, and consequently improve significantly the user experience. One could also leverage on regional \emph{ISP and Private Data centers (E1)} to manage high throughput of write operations (and inter-player transactions) to enable trading objects. Some trades could actually be achieved by having transaction executed directly between the \emph{Tablets \& Mobile Devices (E6)} of players and synchronizing operations towards the \emph{Cloud (E0)} later. Notice that, as discussed previously, enabling users to interact with the application while off-line could be a significant improvement. These could include operations such as item management or even posting trades. Such operations would be verified (and become permanent) when the user device regained connectivity.

Special game features provided by third party businesses could be supported by \emph{Private servers (E4)} being accessed through local networks (supported by \emph{Network Devices (E3)}) located on business premises. Some examples of game features that could be provided this way, considering our case study, include chat rooms for users to meet and interact, distribution of special items related to the location, or special clues for quests.

\paragraph{Intelligent Health Care Services} Consider an integrated and intelligent medical service that inter-connects patients, physicians (in hospitals and treatment centers), and emergency response services\,\footnote{A significative evolution of the Denmark Medical System briefly described in~\cite{fmke}.}, that can leverage on wearable devices (e.g., smart watches or medical sensors), among other IoT devices (e.g., smart pills dispensers), to provide improved and more efficient health care services including, faster handling of medical emergencies, smaller waiting queues for treatment, and tracking health information in the scope of a city, region, or country.

These systems are not a reality today due to, in our opinion, two main factors. The first is the large amounts of data produced by a large number of health monitors, the second is related with privacy issues regarding the medical data of individual patients. Edge computing and the clever usage of different edge resources located in different levels (as discussed previously) can assist in realizing such application. In particular, \emph{Wearables and medical sensors (E7)} can cooperate among them and interact with users' \emph{Mobile Devices (E6)} and \emph{Laptops (E5)}, which can archive and perform simple analysis over gathered data. The analysis of data in these levels could trigger alerts, to notify the user to take medicine, to report unexpected indicators, or to contact emergency medical services if needed. This data could be protected using adequate cryptography primitives and uploaded to \emph{Private Servers (E4)} of hospitals, so that physicians could follow their patients' conditions. Additionally, health indicators aggregates could be anonymously uploaded to \emph{Private Data Centers (E1)} for further processing, enabling monitoring the health quality at the level of cities, regions, or countries to identify pandemics or to co-related them with environmental aspects in these areas. Such a system, if constructed using the right tools and abstractions, and if correctly operated, could have a significant positive impact in the quality of life of many users.

%Budget in pages: 1
\section{Research Challenges}\label{sec:requirements}

%We now discuss concrete research challenges, from a systems perspective, that have to be surpassed in order to fully utilize the potential of edge computing and to realize the case studies discussed previously. We further relate these challenges with previous research that tackled similar challenges and that define a departure point for the realization of this research agenda.

The presented case studies ($\S$\ref{sec:usecases}) rely on the use of multiple edge levels as discussed previously ($\S$\ref{sec:edge}). Other novel edge-enabled applications will have similar requirements. Some of the most challenging aspects of the edge can be found on the high heterogeneity in terms of capacity, location, and management domains of different edge resources and that one has to deal with the increasing number of devices concurrently. We identify the following main challenges to fully realize the potential of edge computing, which we also discuss in relation to our envisioned case studies.

%\paragraph{Resource management:}
\subsection{Managing Edge Resources}
Edge-enable applications will make use of the available resources across different edge levels. To do so, resource management solutions that keep track and manage the high number of computational resources that reside at multiple edge levels are needed. Considering the use case of a \emph{mobile interactive multiplayer game} this translates in two complementary aspects. The first is to track computational resources to enable executing (different) application logic components in cloud platforms, ISP and private data centers, or private servers located in shops. The second is related with mobile devices in close vicinity finding each other and interacting, either directly (using the widely available wireless communication technology such has WiFi Direct or Bluetooth) or through 5G towers.

Todays cloud-based applications are managed by solutions such as Zookeeper\,\cite{zookeeper}, Mesosphere\,\cite{mesosphere}, Yarn\,\cite{yarn}, among others. These often rely on centralized and coordination heavy components which limits the ability of systems that use these building blocks for management to extend towards the edge.
On the other hand, large-scale decentralized resource tracking and management has been previously addressed in the context of decentralized peer-to-peer systems\,\cite{p2p}. Overlay networks\,\cite{leitao:phd} have been used to enable the tracking and communication/coordination among large numbers of resources\,\cite{hyparview,plumtree,cyclon}, efficient application-level routing\,\cite{chord,pastry}, and also decentralized and distributed scheduling of tasks\,\cite{kelips}.
However, most of these solutions assume resources to be homogeneous in both capacity and connectivity which makes them unsuitable for edge environments. Additionally, centralized solutions in this context might not entirely match the hierarchical nature of hybrid cloud-edge systems, that in some sense has been captured by our edge spectrum ($\S$\ref{sec:edge}). %This should be written in a more "providing a hint that it is in this direction"

\subsection{Executing Computations}
Edge-enabled applications require computations to be executed across heterogeneous edge resources located in different levels. Computations cannot be executed in arbitrary edge resources (i.e., any edge level), and shipping computations across different edge levels must cope with heterogeneous execution environments (e.g., virtualization, containers, middleware, different operating systems, etc) and available computational capacity (e.g.,  sensors in E7 cannot perform the same computational tasks as servers in E1). This is illustrated in the \emph{intelligent heath care services} use case, where it is relevant to allow computations that perform initial analysis of data gathered by wearable sensors to migrate between patients mobile devices and laptops according to their availability and granularity of the computation. For instance, these decisions should consider aspects such as: $i)$ computations that reduce the output significantly in relation to the input (summarization, aggregation, filtering) should be executed as closer to data sources as possible to avoid wasting bandwidth from transferring irrelevant data; $ii)$ some computations require access to larger collections of data, and hence should strive to execute in close vicinity to the locations where such data can be accessed.

Research from the software systems community on osmotic computing has already explored this venue, by proposing the migration of microservices, that encode fractions of the computational logic of applications, between cloud infrastructures and the edge according to evolving workloads\,\cite{osmoticcomputing}. Mobile agents\,\cite{mobileagents} and mobile code solutions\,\cite{mobilecode} are proposals that allow having arbitrary code move and execute along a (logical) network of components. However, none of these approaches deal with resource heterogeneity both in terms of execution environment and computational capacity. Amazon\,\cite{lambda@edge} and Google\,\cite{googleedgearch} have enriched their cloud infrastructure to pre-process and redirect HTTP requests to data centers closer to clients. Yet, these only operate at the E0 edge level and extending them towards the edge is an open challenge.
There is a complementary challenge related with this aspect, which is the way applications are designed and implement might have to suffer some additional changes. In addition to reasoning about the application as a set of independent modules (similar to what is done in microservices), such modules might need to be further annotated for instance, to encode restrictions in relation to edge levels where they can execute, or to identify the type of computations performed by them. This can be relevant to allow automated mechanisms to manage the deployment (e.g., replication) and life cycle of different components.

\subsection{Application State Replication}
Edge-enabled applications will naturally need to perform computations over application data; however, as the computations can be scattered through multiple edge levels, and to avoid continuous and potentially high communication overheads with the core of network (in E0), application state should be able to move towards the edge.
This brings additional challenges since not all edge resources can hold the same amount of application data, which motivates the need for effective partial replication solutions. Furthermore, it is imperative that replication protocols manage the dynamic spawn and decommission of partial replicas and provide adequate consistency guarantees, allowing developers to easily reason about their applications.
This is motivated, for instance, by the \emph{mobile interactive multiplayer game} use case, where 5G towers can provide better quality of service for direct interactions among users if these replicate a fraction of the application state for users in their vicinity (i.e., a 5G tower can selectively replicate only the state being accessed and manipulated by users directly connected to it or to neighboring towers). However, since users move, it is essential to be able to freely spawn and remove (to minimize operational costs) partial replicas of the application state in a highly reactive manner. This requires not only the replication unit to be small, but also replication protocols to easily adapt to variations in the number and locations of replicas. Both of these requirements are not found in most replicated data management systems.

There have been many recent works exploring geo-replication, where replicas are dispersed in remote locations. Some focus on offering strong consistency guarantees\,\cite{blotter} while others focus on providing causal+ consistency as to ensure availability\,\cite{cops,eiger,chainreaction,Zawirski15Write}. However, few solutions exploit the use of partial replication. The ones that do so, such as Saturn \cite{saturn}, C3\,\cite{c3}, and Kronos \cite{kronos}, are limited in terms of scalability and inadequate to cope with large number of replicas managed dynamically. This is caused by solutions either relying on centralized components, or employing replication mechanisms that depend either on static (and fragile) control structures to ensure correctness (e.g., statically configured spanning trees connecting all replicas), or control metadata that grows linearly with the number of replicas (e.g., vector clocks).
There has also been few works exploring the combination of multiple consistency models. Gemini\,\cite{redblue} and Indigo\,\cite{indigo} do so per operation type over the data store instead of per replica. Nevertheless, none of these approaches entirely address the data management requirements for future edge-enabled applications.

\subsection{Distributed monitoring}
Dynamically migrating computations and storage components along the edge spectrum requires knowledge about available edge resources, their free capacity, and the current workload (and sources of requests) to which the application is being subjected. This knowledge can be attained by distributed monitoring systems that gather information regarding applications' operation and the load of edge resources. This knowledge can then be employed to perform adequate management decisions. In the \emph{intelligent health care services} use case, choosing the best device to conduct the preliminary analysis of data gathered by wearable sensors requires having (somewhat) up-to-date information regarding the status of each device. Namely, if the device is active, if it has available CPU and RAM to perform the computations efficiently, and its current battery level (or if the device is currently connected to a power source).

There are several previous works that focus on decentralized monitoring of large-scale platforms, typically on cloud and grid infrastructures\,\cite{monitor1,astrolabe,Leitao:danms08}. Many of these solutions resort to gossip-based dissemination protocols\,\cite{plumtree,bimodal} to propagate relevant information towards special sink nodes.
While these solutions offer an interesting departing point for devising new monitoring schemes, they do not consider the monitoring of heterogeneous resources organized in hierarchies as captured in our edge resource spectrum. This means that monitoring information should flow among different resources following the hierarchy, potentially being summarized after a given horizon of the location where the data was generated. This is important to allow local management decisions and avoid central points of failure.

\subsection{Managing Edge Applications}
The management of the life cycle and interactions of large numbers of edge components operating at resources scattered throughout multiple edge levels will be unfeasible by hand, particularly when considering the need to timely adapt the operation of the system in reaction to unpredictable failures of components or sudden surges in access patterns (i.e., peak loads). Considering the use of 5G towers in the \emph{mobile interactive multiplayer game}, spawning application computational components and partial replicas of the game state is manually unfeasible in a timely way considering the potentially large number of available 5G towers. Such process requires autonomic control that, based on monitoring information, can automatically trigger application management mechanisms.

Autonomic systems have been proposed and studied since 2001\,\cite{autonomic} and have been applied to multiple types of systems from web services\,\cite{autonomics09}, to management of grid resources and others\,\cite{autonomics}. Typically, autonomic systems are designed around the $MAPE$-$K$ architectural design, where the system has $M$onitoring, $A$nalysis, $P$lanning, and $E$xecution components that are interconnected by a common $K$nowledge base. The main challenge in making edge-enabled applications autonomic is due to the typical central nature of the Planning (and potentially Analysis and Execution) components. Large-scale edge-enabled applications combining multiple components scattered across several edge levels will require decentralized planning schemes, capable of operating (and executing) reconfigurations of the system with incomplete and partial knowledge. While some previous works have explored this\,\cite{decentralizedautonomic}, they are still far from meeting the needs of future edge-enabled applications.

\subsection{Security \& Data protection}
Future edge-enabled applications will manipulate sensitive user data. Doing so can compromise users' privacy. Furthermore, executing computations and storing data in hardware controlled by individual users can also compromise data integrity. In the \emph{intelligent health care services} use case, medical data would be stored in user devices and in private servers in hospitals. This data is highly sensitive as it can easily compromise the privacy of patients, while compromising its integrity could have serious medical implications, for instance, if a patient receives the incorrect treatment. To address this, data protection schemes must be employed in future edge applications, this is typically achieved by using cryptography primitives.

Homomorphic\,\cite{homomorphic} and partially homomorphic\,\cite{pallier, elgamal} encryption schemes allow computations to be performed in the encrypted domain. This is beneficial as it allows sensitive data to never be exposed to untrusted parties, while still allowing such parties to cooperate in the management of that data. However, these solutions have high computational cost and produce high cipher-text expansion, meaning that these solutions are impractical in many edge devices and would further increase the pressure put on network links, being therefore unsuitable to support edge computations.
A more promising alternative is to use schemes based on (efficient) symmetric cryptography, enabling some operations (such as indexing and search) in the encrypted domain. This has been demonstrated for a few data formats\,\cite{ferreira1,ferreira2}. Nevertheless, these schemes do not provide guarantees of data integrity.
Protecting data integrity can rely on trusted hardware\,\cite{trustedhardware} (e.g., IntelSGX\,\cite{sgx}) that provide attestation and verification mechanisms for outsourced computations. Unfortunately, limited resources and complex key management/distribution schemes, makes the use of trusted hardware an open challenge today.

\section{Conclusion}\label{sec:agenda}
In this section we discuss the research agenda (focused on a computer systems perspective) that has to be pursed by the community to enable: $i)$ the sustainable growth of edge computing and applications leveraging such architectures; and $ii)$ to realize our vision of future cloud/edge hybrid architectures and novel edge-enabled applications as the ones discussed previously in this paper.
We note that our vision is complementary to recent efforts in devising reference architectures and exploitation plans for edge computing, including academic efforts\,\cite{humancentric} as well as those of the OpenFog Consortium, which have proposed a Fog reference architecture\,\cite{openfog}, and ETSI to standardize Mobile (or Multi-Access) Edge computing architectures\,\cite{etsi}.

%\begin{description}
	\textbf{Scalable Hierarchical Resource Management:}
		As we discussed  previously, novel edge-enabled applications will tend to make use of an increasing number of edge resources (across different levels of the edge spectrum). This will lead to more complex deployments, composed of larger number of application components executing in diverse resources. To allow the growth of edge applications, novel solutions to manage edge resources will be required. Centralized solutions will have to be abandoned in favor of decentralized and inherently scalable solutions. Ideally, such solutions will take advantage of the hierarchical nature of the edge spectrum. While existing approaches do not fit with this model, we believe that an interesting departure point can be found in peer-to-peer overlays. These are highly decentralized, robust, and can easily be adapted to promote hierarchical topologies. This however, will require fundamental research to devise novel mechanisms to enable both the construction of such overlays, while also taking into consideration the heterogeneity of edge resources, as well as devise adequate fault-tolerance mechanisms (it has been shown that biasing overlay topologies can compromise their robustness\,\cite{xbot}). In addition, monitoring solutions will have to be devised to operate efficiently over such topologies.

	\textbf{Enable Computations to Move:}
	To ensure that novel edge-enabled applications are also sustainable from an economic point of view, their operation will have to be agile and capable of adapting to evolving conditions. This implies that different components of the application logic might be deployed across different levels of the edge spectrum and with different number of copies. Such deployments will have to take into consideration several factors, being the more obvious: $i)$ edge resource availability, $ii)$ edge resource cost, $iii)$ user location, and $iv)$ user load. This implies that parts of the application logic must be able to be migrated and replicated (or decommissioned) very effectively. Moreover, these components must be able to inform the runtime of some specifications (or requirements) of their operation to avoid incorrect deployment decisions. This will require a better understanding of the different types of computations that exist and their fundamental (and distinguishable) properties. For instance, we have already pointed out two potential classes: \emph{filtering} and \emph{aggregation and summarization}. It is well known that filtering is an operation that can operate over individual data records, and different instances of filtering computations can operate completely independently. Additional research efforts are required to map and fully understand other computations. Finally, mechanisms to automatically classify such portions of application logic would significantly contribute to avoid incorrect deployment decisions at runtime.

	\textbf{Dynamic and Partial Data Replication:}
	It is expected that novel and emerging edge-enable applications will manipulate ever increasing amounts of data. All of that data has to be stored as to become accessible and enable efficient operation. Resorting to large data storage management systems operating in the cloud and featuring full replication (across different data centers) is not viable to support applications with large number of components operating in wide disperse areas. Instead, new and fundamental research has to be conducted to evolve existing distributed storage solutions so that they can organically scale from data centers to the edge, in particular by enabling data to use replication units significantly smaller than current solutions. This is important because small edge devices might only be able to replicate a few hundreds data objects for very particular users. This design shift will require also the design and implementation of effective and efficient replication algorithms that provide both sensible consistency guarantees (such as Causal consistency) while being able to scale to an unprecedented number of replicas. Existing research on scalable partial replication solutions providing causal+ consistency\,\cite{saturn,c3} are interesting starting points to lead the research on this vector.

	\textbf{Decentralized Autonomic Systems:}
	Future edge-enabled applications will impose high burden on application operators due to the large-scale and high number of components that will, together, contribute to the sustained and efficient operation of these applications. Manual management will quickly become unfeasible. To address this challenge autonomic and decentralized management mechanisms will have to be put into place. While autonomic control has found significant success in cloud-based infrastructures, for instance with features such as elasticity, these are fundamentally centralized and depend on complex monitoring infrastructures. Additional research will be required to fully realize the potential of autonomic computing, namely by enabling its operation in a decentralized and hierarchical way. In this context, decentralized and hierarchical resource management mechanisms (discussed above) will play an important role, but will have to be enriched with planning and locally coordinated executors. Machine learning mechanisms may play an interesting role to allow such autonomic management mechanisms to effectively identify adequate deployment configurations however, additional research is required to also decentralize such mechanisms.

	\textbf{Security \& Data protection:}
	Security and data protection are essential aspects of any large-scale system in operation nowadays. However, as applications extend beyond the (safe) boundaries of cloud infrastructures new challenges naturally emerge that cannot be easily tackled by current security solutions. User data that is stored (even if temporarily) on many devices and different administrative domain will have to be protected, especially if it contains sensitive information. However, to avoid limiting the functionality provided by novel edge-enabled applications, simple encryption is not enough. This implies researching novel mechanisms to allow operations to be performed in the encrypted domain. This can be achieved by either	overcoming practical limitations of homomorphic and partially-homomorphic encryption schemes, or by expanding the functionality proposed by recent works on searchable symmetric encryption. Additionally, protecting the data from unauthorized access or incorrect manipulation requires significant efforts in devising novel decentralized access control mechanisms and enforcement mechanisms, that can operate across all edge levels without significant performance penalties. One interesting departure point for addressing issues related with data integrity is the use of trusted hardware platforms, particularly in devising novel verification mechanisms that operate in a machine-to-machine way, creating verifiable traceability over transformations executed on application data.

\bibliographystyle{plain}
\bibliography{bib}

\end{document}